\documentstyle [11pt,psfig]{article}
\def\a{\alpha}    \def\b{\beta}    \def\g{\gamma}    
\def\e{\epsilon}      \def\t{\theta}    \def\l{\lambda}
       \def\O{\Omega}
\def\p{\partial} 
\def\be{\begin{equation}}
\def\ee{\end{equation}}
\def\bea{\begin{eqnarray}}
\def\eea{\end{eqnarray}}
\def\bean{\begin{eqnarray*}}
\def\eean{\end{eqnarray*}}
\def\r#1{(\ref{#1})} 
\def\la#1{\label{#1}}                                     
\def\c#1{\cite{#1}}

\begin{document}
\rightline{September 1997}
\vskip 1 true cm                                                               
{\Large   \centerline{The Camassa--Holm Equation:} 
          \centerline{A Loop Group Approach} 
          \vskip 1 true cm
          \centerline{Jeremy Schiff}             }
{\small   \vskip 0.5 true cm
          \centerline{Department of Mathematics and Computer Science}
          \centerline{Bar--Ilan University, Ramat Gan 52900, Israel}    
          \centerline{e-mail: schiff@math.biu.ac.il}      }
\vskip 1 true cm

\noindent {\bf Abstract.} 
A map is presented that associates with each element of a loop group a
solution of an equation related by a simple change of coordinates to 
the Camassa--Holm (CH) Equation. Certain simple automorphisms of the
loop group give rise to B\"acklund transformations of the equation.
These are used to find 2-soliton solutions of the CH equation, as well as
some novel singular solutions. 

\vskip 1 true cm

\section{Introduction}

Substantial interest is accumulating in the Camassa--Holm (CH) equation:
\be
u_t = 2f_xu+fu_x\ , \qquad  u = {\textstyle{\frac12}}f_{xx}-2f\ . \la{ch}
\ee
This equation has been believed to be integrable for many years \c{Fuch1,FF},
but only recently has it been widely studied, following the work of Camassa
and Holm \cite{CH} showing that it describes shallow water waves. Camassa 
and Holm found that this equation exhibits ``peakons'', i.e. solitary wave
solutions with discontinuous first derivative at their crest (here $f$ is 
regarded as the fundamental field; for peakon solutions $u$ is just a moving 
delta function). Multipeakon solutions can be found, and are related to an 
integrable finite dimensional Hamiltonian system, which has been exhaustively
studied \c{Calogero,RB}. Both numerical and analytic studies
\c{CHH,CE} suggest that for suitable initial data \r{ch} describes
the decomposition of the initial data into peakon components; in particular
for analytic initial data with $u$ of mixed sign, the first derivative of $f$
develops a discontinuity in finite time. In addition to peakon solutions,
\r{ch} is  known to have analytic soliton solutions tending to finite 
nonzero depth at spatial infinity \c{Fokas,LO}; these converge to peakon 
solutions as the depth at infinity tends to zero. 

A number of papers have appeared explaining various aspects of the integrable 
structure of CH \c{FOR,Fuch2,OR,Ros2,me1}. Despite these results, much 
work remains to be done, particular in regard to generating explicit 
solutions. The aim of the current paper is to present the analog, 
for CH, of a cornerstone of KdV theory, the {\em Segal-Wilson map} 
\c{SW}. The Segal-Wilson map associates with each element of a loop group 
a solution (possibly with singularities) of KdV. In \c{Wilson}, Wilson gave a 
very explicit version of this map, writing down a huge class of solutions of 
the modified KdV equation. Here I will give an analogous formula for CH;
more precisely, I give a map from a loop group to the space of solutions
of an equation related by a simple change of coordinates to the CH equation,
which I will call the associated Camassa-Holm (ACH) equation.
The application I will make of this result is the construction of 
B\"acklund transformations (BTs) for the ACH equation. In the case of the KdV
equation, it is known that BTs have their origins in simple automorphisms of 
the relevant loop group \c{me2}, and by looking at similar transformations 
here, BTs can be derived for ACH. These facilitate the construction of new
solutions of CH, along with a new formula for 2-soliton solutions. (A formula
for 2-soliton solutions has already been given in \c{Alber}, using a
different approach, with which I will not compare here.)

This paper is structured in a logically incorrect fashion, but one which
I hope will enable others working on the CH equation to read the results
obtained without going into details of the loop group construction. 
Section 2 contains all the results that do not require some understanding
of loop groups: Here I define the ACH equation, explore
its elementary solutions and properties, give two BTs for ACH
(derived later by loop group techniques), and use the BTs
to study less elementary solutions of ACH, and the corresponding
solutions of CH. Section 3 contains the details of the map from a loop
group to solutions of ACH, the main result of the paper.
Finally, Section 4 contains the  derivation of the two BTs using loop 
group methods, which logically should precede much of the material
in section 2.

The reader will see  in section 2 that this paper only studies
solutions of CH for which $u$ is of constant sign (only then is the 
transformation to ACH defined). I have tried hard to find a loop group
construction that gives rise  to mixed sign solutions of CH, but
without success. This, along with the results of \c{CE},
leads me to conjecture \r{ch} is in some sense ``more''
integrable for solutions of constant sign than for solutions of mixed
sign (in \c{CE} it is shown that if a solution of \r{ch}
is of constant sign at some time, then it remains so). The exact sense of
this remains to be clarified, but it is certainly clear that the CH equation
presents an interesting challenge to the integrable systems community.

A few more introductory points: First, in the current paper I limit myself
to the study of \r{ch}, and not the related equation obtained by replacing
the definition of $u$ in \r{ch} by $u=\frac12 f_{xx}+2f$, which admits
a compacton solution \c{LO,Ros1,RH}. Second, note the choice of 
coefficients I have made in \r{ch} differs slightly from that in \c{CH};
in particular for my choice of coefficients positive elevation peakons
move to the left. Third, I note that the change of coordinates from CH
to ACH is suggested in \c{Fuch2}. And finally, I draw the reader's 
attention to the papers \c{CM}, which study the periodic problem for
the CH equation.

\section{The ACH Equation and its B\"acklund Transformations}

\noindent{\em Proposition 1.} There exists a one to one correspondence 
between $C^\infty$ solutions of \r{ch} with $u$ positive and 
$C^\infty$ solutions of 
\be 
\dot{p} = p^2f', \qquad f=\frac{p}4\left(\frac{\dot{p}}{p}
   \right)'-\frac{p^2}2\ ,\la{ch2}
\ee
with $p$ positive. Here $p,f$ are functions of $t_0,t_1$, 
a prime denotes differentiation with respect to $t_0$, and a dot
differentiation with respect to $t_1$. Equation \r{ch2} will be referred
to as the {\em Associated Camassa-Holm} (ACH) Equation.

\smallskip

\noindent{\em Proof.} Suppose we have a solution to \r{ch} with 
$u$ positive, and let $p=\sqrt{u}$. Then
$p_t=(pf)_x$. It follows that we can define a new set of coordinates
$t_0,t_1$ (the reason for this notation will become clear in section 3) 
via  \be
dt_0 = p dx + pf dt\ , \qquad dt_1 = dt\ . \la{co-ch}
\ee
In the new coordinates equation \r{ch} becomes \r{ch2}. To go from a 
solution of \r{ch2} to a solution of \r{ch}, we note that the change of
coordinates \r{co-ch} implies
$$
\pmatrix{ \frac{\p t_0}{\p x} & \frac{\p t_0}{\p t} \cr
          \frac{\p t_1}{\p x} & \frac{\p t_1}{\p t} \cr}
= \pmatrix{ p & pf \cr
            0 & 1  \cr},
$$
and so
$$ 
\pmatrix{ \frac{\p x}{\p t_0} & \frac{\p x}{\p t_1} \cr
          \frac{\p t}{\p t_0} & \frac{\p t}{\p t_1} \cr}
= \pmatrix{ \frac1{p} & -f \cr
            0 & 1  \cr}.
$$
Given a solution $p,f$ of \r{ch2}, with $p$ non-vanishing,
we find $x$ as a function of $t_0,t_1$ by integrating
\be \frac{\p x}{\p t_0}=\frac1{p}\ ,
\qquad \frac{\p x}{\p t_1}=-f\ .
\la{xfts} \ee
By the first equation of \r{ch2}, these equations are integrable. 
We  clearly can identify $t$ and $t_1$; and since $\frac{\p x}{\p t_0}>0$
it follows that the map between $x$ and $t_0$ for fixed $t=t_1$ is
one to one (and $C^\infty$). Thus we can express $t_0,t_1$ in terms of 
$x,t$ to obtain a solution of \r{ch} (with $u=p^2$). 
It is clear that this sets up a one to one correspondence between 
``positive'' solutions of the two equations.

\smallskip

\noindent{\em Note.} Clearly a solution of \r{ch2} with $p$ negative also
gives rise to a solution of \r{ch}. A solution of \r{ch2} for which $p$ 
has zeros will in general give rise to a number of solutions of \r{ch}:
in integrating \r{xfts} we obtain a relationship between $t_0$ and $x$
(for fixed $t_1=t$) which is many to one. We will see an example of this
below. In the other direction, a solution of \r{ch} with $u$ always negative
can be used to give a solution with $u$ always positive by the replacements
$u\rightarrow -u$, $f\rightarrow -f$, $t\rightarrow -t$, and from this
we can obtain a solution of \r{ch2}. But
there is no apparent way to obtain a 
solution of \r{ch2} from a solution of \r{ch} in which $u$ is allowed to 
change sign. Finally, the correspondence also extends to solutions 
with point singularities; below we will see  an example of this too.

\smallskip

\noindent{\em Proposition 2.} $p(x,t)=\phi(x-ct)$ ($c\not=0$) 
solves \r{ch2} if 
\be
(\phi')^2 = -\frac4{c}\phi^3 + \a\phi^2+\b\phi+4\ , \la{sol}
\ee
where $\a,\b$ are arbitrary constants. In particular we have 
``soliton solutions''
\be 
\phi(z) = A\ {\rm sech}^2 \left(\sqrt{\frac{A}{c}}\ z\right) + h\ , \qquad
    A = \frac{c}{h^2}-h\ , \la{solsol}
\ee
(here $h\not=0$ and $h^3\ {\rm sgn}(c)<\vert c\vert$), and 
(singular) ``rational solutions'' 
\be \phi(z)=c^{1/3} - \frac{c}{z^2}\ .  \la{ratsol}\ee

\smallskip

\noindent{\em Proof.} This is a straightforward computation.
Equation \r{sol} is familiar from the 
theory of the KdV equation, but the nonzero constant term implies that 
if  $\phi'\rightarrow 0$ at spatial infinity, then $\phi$ cannot 
go to zero there, as seen in the forms of both the soliton and rational
solutions. In the soliton solutions the parameter
$h$ is the asymptotic height; for the rational solutions the asymptotic
height is $c^{1/3}$, a limiting case of the heights allowed for soliton
solutions.

\smallskip

\noindent{\em Corresponding Solutions of CH.} It is a straightforward
but arduous matter to translate the solutions of ACH just presented
to solutions of CH, following the procedure in the proof of
proposition 1. From the soliton solutions of ACH
we obtain the soliton solutions of CH, which for $c>0$ take the form
\bea
f &=& -\frac{h^2}2 + c\left(\frac1{A\ {\rm sech}^2X+h}-\frac1{h} \right) 
      \ , \nonumber\\
&& \\
u &=& (A\ {\rm sech}^2X+h)^2\ ,  \nonumber
\eea
where $A=(c/h^2)-h\ $, and $X~\left(=\sqrt{A/c}(t_0-ct_1)\right)$ 
is determined from 
\be
x-\left(\frac{c}{h}+\frac{h^2}2\right)t =
\sqrt{\frac1{1-h^3/c}}\ X + \frac12\ln\left(\frac{me^{2X}+1}{e^{2X}+m}
\right),\qquad m=\left(\sqrt{\frac{c}{h^3}} - 
\sqrt{\frac{c}{h^3}-1} \right)^2\ .    \nonumber\ee
The speed of the solution as a solution of CH is 
$\tilde{c}=c/h+h^2/2$, differing from the speed $c$ of the solution as a 
solution of ACH. To understand the nature of this solution, it is 
useful to look at the relation of $x-\tilde{c}t$ and $X$ in the limits
$X\rightarrow\pm\infty$ and $X\rightarrow 0$: For $X\rightarrow\pm\infty$
$$ x-\tilde{c}t = \sqrt{\frac1{1-h^3/c}}\ X \pm
       \ln\left(\sqrt{\frac{c}{h^3}} - \sqrt{\frac{c}{h^3}-1} \right)
      + o(1)\ , $$
and for $X\rightarrow 0$
$$ x-\tilde{c}t = \left( \sqrt{\frac1{1-h^3/c}} - \sqrt{1-h^3/c}  
                   \right) X  + o(X)\ . $$
In figure 1, a plot of $x-\tilde{c}t$ against $X$ is given for the value
$h^3/c=1/3$, as well as plots of $f/h^2$ and $u/h^2$ against $x-\tilde{c}t$
for this value. The limit $h\rightarrow 0$ with $\tilde{c}$ constant is
--- according to \c{LO} --- the peakon limit (or, rather, since we are
looking at right moving solutions, the anti-peakon limit); I leave it
as an interesting exercise to the reader 
to show that in this limit we indeed obtain the 
anti-peakon solution
\be f=-\tilde{c}\exp(-2\vert x-\tilde{c}t\vert)\ , \qquad
    u=2\tilde{c}\delta(\vert x-\tilde{c}t\vert)\ . \ee

\begin{figure}
\centerline{\psfig{figure=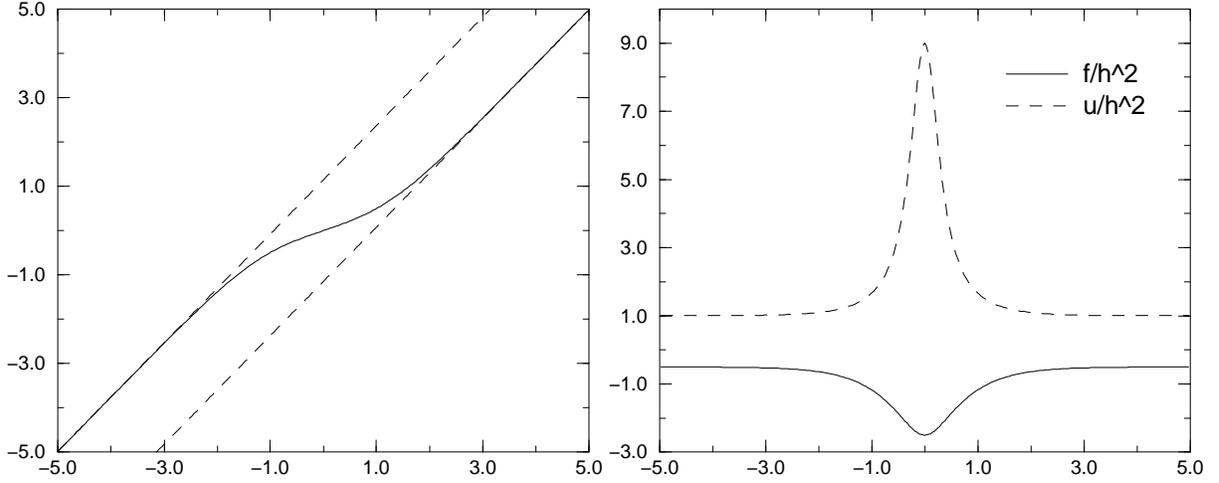,height=2.5in}}
\caption{The soliton solution, for $h^3/c=1/3$. On the left, a plot of $x
-\tilde{c}t$ versus $X$ is displayed; the straight lines show the $X
\rightarrow\pm\infty$ behavior. On the right the functions $u/h^2$ and 
$f/h^2$ are plotted against $x-\tilde{c}t$.}
\end{figure}

Turning  now to the rational solutions \r{ratsol}, for these $\phi$ 
changes sign twice (at $z=\pm c^{1/3}$) and has a singularity (at $z=0$).
Following the method of translation back to solutions of CH
Leads to
\be f = \tilde{c}\ \frac{1-\frac13 X^2}{X^2-1}\ , \qquad
    u = \frac23\tilde{c}\left(1-\frac1{X^2}\right)^2\ , \ee
where $X~\left(=zc^{-1/3}\right)$ is determined by
\be x-\tilde{c}t=X+\frac12\ln\left\vert\frac{X-1}{X+1}\right\vert\ , \ee
and $\tilde{c}=\frac32c^{2/3}$. The map from $X$ to 
$x-\tilde{c}t$ is three to one,
so  here a single solution of ACH gives three solutions of CH,
corresponding to the ranges on which $\phi$ is of definite sign,
viz. $X<-1$, $-1<X<1$ and $X>1$. In the middle range the solution
has a singularity at $X=0$. The three solutions are illustrated in figure 2. 
For $X<-1$ and $X>1$, the solutions for $u$ take the form of a
``kink'', with the limit at one end being approached polynomially
and at the other end exponentially; the solution for $f$ is finite at one end,
and diverges exponentially at the other end. 
For the range $-1<X<1$, the solution for
$u$ is a ``spikon'' (an infinitely peaked soliton), and  $f$ has a cusp at 
$x-\tilde{c}t=0$ and exponentially diverges at infinity. 
(For $x-\tilde{c}t\approx 0$ we find 
$f\approx-1-2\cdot 3^{-1/3}(x-\tilde{c}t)^{2/3}$, and 
there is a cusp at $x-\tilde{c}t=0$, not a simple corner as appears from 
the low resolution plot in figure 2.) One might hope to ``splice''
together the two solutions for $X<-1$ and $X>1$ 
to form a finite height peakon solution for $f$ with polynomial decay at 
infinity; this does not seem to be possible.

\begin{figure}
\centerline{\psfig{figure=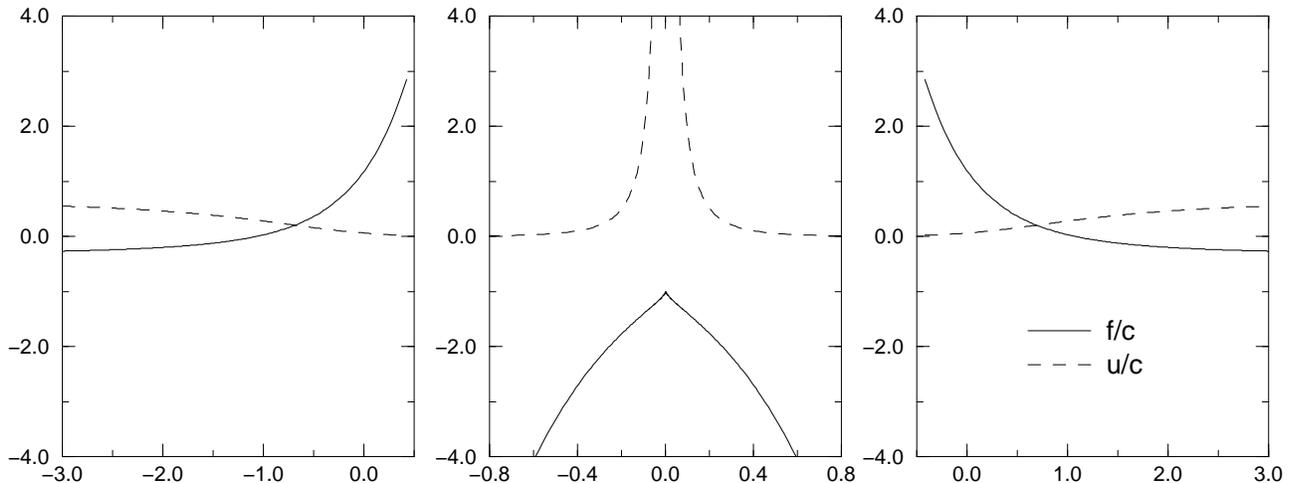,height=2.5in}}
\caption{The solutions of CH obtained from 
rational solution of ACH, for the ranges $X<-1$, $-1<X<1$ and
and $X>1$ (from left to right).}
\end{figure}

\noindent{\em Note.} I have not, as of yet, explored the solutions 
of \r{ch} corresponding to the cnoidal wave solutions of \r{ch2}.

\smallskip

\noindent The great advantage of ACH over CH is that ACH has standard
B\"acklund transformations. These are presented here (and can be verified
by direct computations), but their derivations will be given in section 4.

\smallskip

\noindent{\em Proposition 3.} The ACH equation \r{ch2} has the B\"acklund
transformation $p\rightarrow p-s'$ where 
\bea
s' &=& -\frac{s^2}{p\t}+p\t\left(\frac1{p^2}+\frac1{\t}\right) \ ,
  \la{bt1}\\
\dot{s} &=& -s^2 + \frac{\dot{p}}{p} s + \t(\t-2f)\ . \la{bt2}
\eea 
This is a strong BT (in the sense that \r{bt1} and \r{bt2} are only
consistent if $p,f$ obey \r{ch2}). $\t$ is a parameter, $\t\not=0$. 
Under the BT
\be
f \rightarrow f - \frac{\dot{s}}{p(p-s')}\ .
\ee

\smallskip

\noindent Applying this BT to the constant solution with $p=h$ and
$f=-h^2/2$, we find we can take
\be
s=h\t\sqrt{\frac1{h^2}+\frac1{\t}}
     \tanh\left(\sqrt{\frac1{h^2}+\frac1{\t}}(t_0+\t ht_1 + C)\right)
\la{s1}\ee
or 
\be
s=h\t\sqrt{\frac1{h^2}+\frac1{\t}}
     \coth\left(\sqrt{\frac1{h^2}+\frac1{\t}}(t_0+\t ht_1 + C)\right),
\la{s2}\ee
where $C$ is a constant of integration. Taking the choice \r{s1} for 
$s$, with $C=0$ and $\t=-c/h$, returns the soliton solution \r{solsol}.
The choice \r{s2} gives a singular solution. Repeated application of 
the BT is simplified by the following nonlinear superposition formula:

\smallskip

\noindent {\em Proposition 4.} If application of the above BT with parameter
$\t_1$ to a solution $p$ of \r{ch2} yields the new solution $p_1=p-s_1'$,
and application with parameter $\t_2$ yields the new solution 
$p_2=p-s_2'$, then 
\be 
p_{12}=
p-\left(\frac{(\t_1-\t_2)(\t_1\t_2-s_1s_2)}{\t_2s_1-\t_1s_2}\right)'
\la{super}\ee
is also a solution of \r{ch2} (arising from repeated application of the 
BT with parameters $\t_1$ and $\t_2$ --- in either order --- to $p$).
The corresponding $f$ takes the form
\be
f_{12}= f- \t_1\t_2(\t_1-\t_2)\left(\frac{p}{p_{12}}\right)
        \left(\frac{\dot{s_1}-\dot{s_2}}{(\t_2s_1-\t_1s_2)^2}\right) \ .
\la{superf}\ee

\smallskip

\noindent{\em Proof (outline only).} 
The formulae \r{super} and \r{superf} are best proven by direct
verification using a symbolic manipulator. As regards their derivation, the
starting point is commutativity of BTs with different parameter values, for
which an argument can be given at the loop group level (see section 4). 
Given this, suppose applying the BT with 
parameter $\t_2$ to $p_1$ gives $p_{12}=p_1-s_{12}'$, and applying the BT
with parameter $\t_1$ to $p_2$ gives $p_{21}=p_2-s_{21}'$. If $p_{12}=
p_{21}$, then $(s_1+s_{12})'=(s_2+s_{21})'$, and this suggests 
looking at the possibility that $s_1+s_{12}=s_2+s_{21}$. Since from the BT
we have expressions for the derivatives of $s_1,s_2,s_{12},s_{21}$, we can
differentiate this relationship to find other algebraic relationships
from which $s_{12}$ and $s_{21}$ can be determined (due to their length,
I do not reproduce these calculations here). This is the origin of 
the formula \r{super}. 

\smallskip

\noindent
2-soliton solutions of ACH are now easily found using the superposition 
formula on the constant solution $p=h$, $f=-h^2/2$. Taking $h>0$, 
$\t_2<\t_1<-h^2$, $s_1$  of the form \r{s1} and $s_2$ 
of the form \r{s2} we obtain
\be
p=h-h(\t_1-\t_2)\cdot  
 \frac
 {(\t_2+h^2) + (\t_1-\t_2)\tanh^2y_2 - (\t_1+h^2)
  \tanh^2y_1\tanh^2y_2}
 {\left(\sqrt{\t_1(\t_2+h^2)}  - \sqrt{\t_2(\t_1+h^2)} 
   \tanh y_1 \tanh y_2\right)^2}\ ,  \la{2sol} 
\ee
where $y_1=\sqrt{1/h^2+1/\t_1}(t_0+\t_1 h t_1 +C_1)$ 
and   $y_2=\sqrt{1/h^2+1/\t_2}(t_0+\t_2 h t_1 +C_2)$. For
$\t_2<\t_1<-h^2$ it is simple to check this is  nonsingular, and moreover 
that $p>h$, so in particular $p$ has no zeros and $f$ is nonsingular 
too. For $f$ we find the formula
\bea
f &=& -\frac{h^2}2 - (\t_1-\t_2)\cdot \la{2solf}\\
 && \frac
 {\t_2(\t_2+h^2) + (\t_1-\t_2)(h^2+\t_1+\t_2)\tanh^2y_2 - \t_1(\t_1+h^2)
  \tanh^2y_1\tanh^2y_2}
 {\left(\sqrt{\t_2(\t_2+h^2)}  - \sqrt{\t_1(\t_1+h^2)} 
   \tanh y_1 \tanh y_2\right)^2
   - (\t_1-\t_2)^2 \tanh^2 y_2}\ . \nonumber 
\eea
To translate these results into  2-soliton solutions of CH  
we follow the procedure of proposition 1: 
$f$ is as given in \r{2solf} and $u=p^2$, where $p$ is as given 
in \r{2sol}, but now $t_1$ must be replaced by $t$, and $t_0$ is
a parameter, related to the coordinates $x,t$ by 
\be 
x = \int_0^{t_0} \frac1{p(t_0',t_1)} dt_0' - \int_0^{t_1} f(0,t_1') dt_1' 
\ee
(which solves \r{xfts}). I cannot see how to evaluate these integrals
analytically, but they can be approximated in various limits, as well
as evaluated numerically. In figure 3, snapshots of a 2-soliton solution
of ACH are given, and in figure 4 the corresponding pictures of the 
corresponding solution of CH are shown. One 
important feature of the passage from ACH to CH is that the speeds of 
the soliton components change; for ACH the speeds are $c_i=-\t_ih$, $i=1,2$,
and for CH they are $\tilde{c}_i=c_i/h+h^2/2=-\t_i+h^2/2$. 

\begin{figure}
\centerline{\psfig{figure=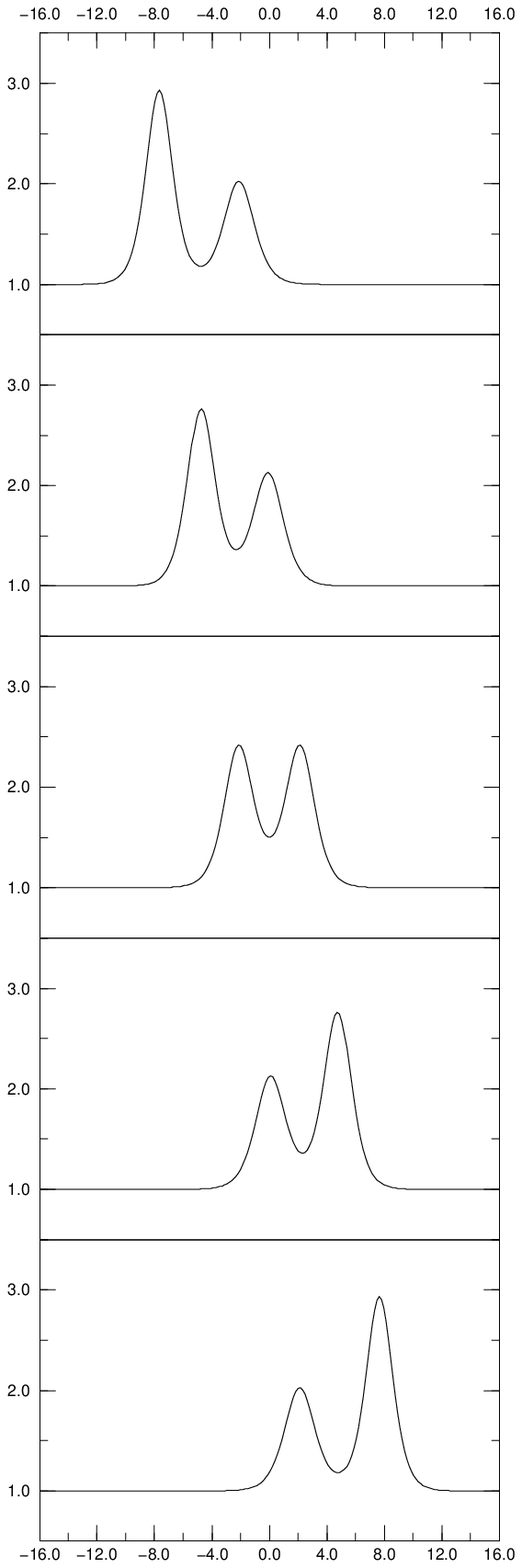,height=6.5in}~~~~~~~~~~~~~~~
            \psfig{figure=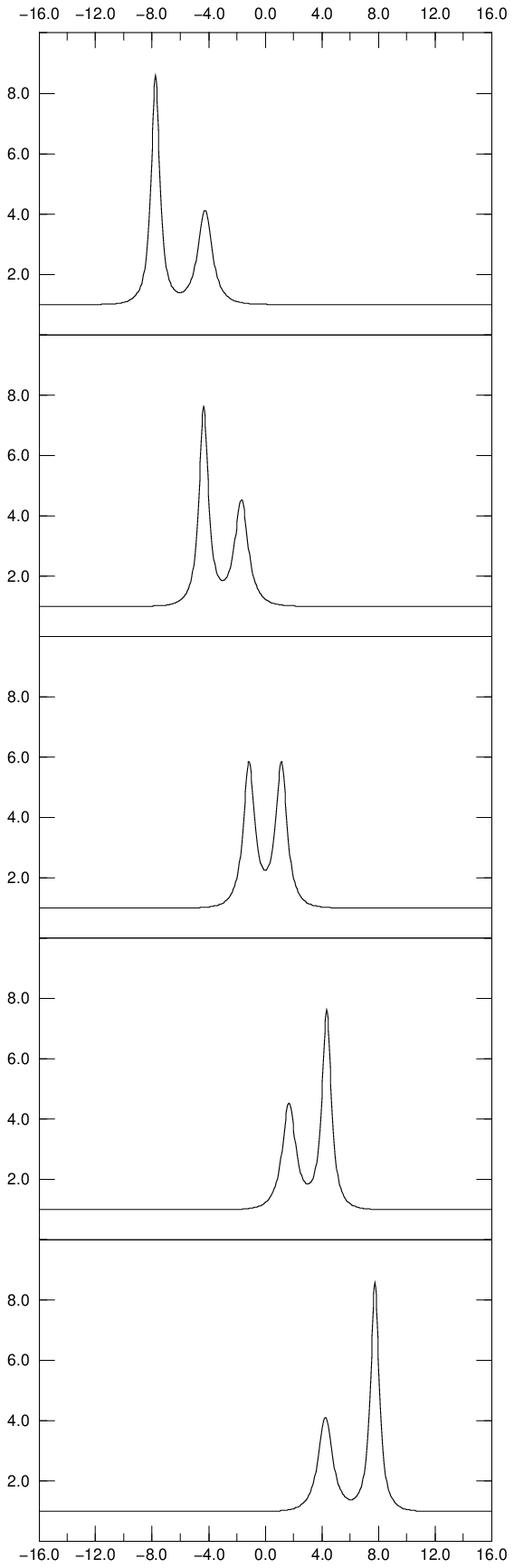,height=6.5in}  }
\smallskip
\smallskip
\noindent From left to right:
\smallskip
\caption{The 2-soliton solution of ACH given by equation \r{2sol}, with
$h=1$, $\t_1=-2$, $\t_2=-3$, $C_1=C_2=0$. Plots are of $p$ as a function
of $t_0$, for $t_1=-3,-2,-1,0,1,2$ (from top to bottom).}
\caption{The corresponding 2-soliton solution of CH . Plots are of 
$u$ as a function of $x$, for $t=-3,-2,-1,0,1,2$ (from top to bottom).}
\end{figure}

I proceed to a second BT for ACH, which will also be derived
later by loop group techniques.

\smallskip

\noindent{\em Proposition 5.} The ACH equation \r{ch2} has the 
strong B\"acklund transformation 
\be p \rightarrow  p\left[ \left( 1-\frac{pBB'}{\tau}\right)^2 - 
                           \frac{B^4}{\tau^2}        \right]\ ,
\ee
where $B$ and $\tau$ satisfy the equations.
\bea
(pB')' &=& B\left(\frac1{p} + \frac{p}{\t} \right) 
\qquad\qquad
\dot{B} ~=~ \left(\frac{-\dot{p}}{2p}\right)B + \t pB'     \la{BT2e1} \\
\tau'  &=& \frac{pB^2}{\t} 
\qquad\qquad~~~~~~~~~~
\dot{\tau} ~=~ \theta( p^2 B'^2 - B^2)           \la{BT2e2}
\eea
and $\t$ is a parameter, $\t\not=0$.

\smallskip

\noindent
This BT looks  more involved than the previous one, but 
it is actually very similar to implement. The apparently
difficult part of implementation is solving the first equation 
of \r{BT2e1}, a second order linear equation determining the $x$-dependence
of $B$. This is, however, just a linearization of the Riccati 
equation \r{bt1} that appears in the previous BT: if $B$ solves the first
equation of \r{BT2e1} then $s=p\theta B'/B$ solves \r{bt1}. 

As an example of the use of this BT, consider its action on the 
trivial solution $p=h$, $f=-h^2/2$ ($h$ constant). One allowed choice
of $B,\tau$ is
\bean
B &=& K_1 \exp\left(\sqrt{\frac1{h^2}+\frac1{\t}}(t_0+h\t t_1)\right) \\
\tau &=& K_1^2\left(\frac{h}{2\t\sqrt{\frac1{h^2}+\frac1{\t}}} 
    \exp\left(2\sqrt{\frac1{h^2}+\frac1{\t}}(t_0+h\t t_1)\right) + K_2
    \right),
\eean
where $K_1,K_2$ are constants,
and it is straightforward to show that this returns the 1-soliton 
solution with speed $-h\t$.
A more general possibility for $B,\tau$ is
\bean
B &=& K_1\cosh\left( \sqrt{\frac1{h^2}+\frac1{\t}} (t_0+h\t t_1+K_2)
\right) \\
\tau &=& \frac{K_1^2 h}{2\t} \left( t_0-\frac{\t}{h}
\left(2\t+h^2\right)t_1+K_3
+ \frac1{2\sqrt{\frac1{h^2}+\frac1{\t}}} \sinh 
\left(2 \sqrt{\frac1{h^2}+\frac1{\t}} (t_0+h\t t_1+K_2)
\right) \right),
\eean
where $K_1,K_2,K_3$ are constants, with resulting solution of ACH 
\be
p = h-4h\left(1+\frac{\t}{h^2}\right)\frac
{\left(\frac{2\t}{h^2}\right)(1+\cosh y_2)+y_1\sinh y_2}
{(y_1+\sinh y_2)^2}\ ,
\la{newach}\ee
where $y_1=2\sqrt{\frac1{h^2}+\frac1{\t}}(t_0-\frac{\t}{h}(2\t+h^2)t_1+K_3)$,
$y_2=2\sqrt{\frac1{h^2}+\frac1{\t}}(t_0+{\t}{h}t_1+K_2)$.
This solution, which is illustrated in figure 5 (for $\t<0$) and
figure 6 (for $\t>0$), should presumably be
considered as the superposition of a soliton and a simple rational
solution. For almost every value of $t_1$, it has a singularity at a 
single value of $t_0$, and two zeros, giving singularities of the 
corresponding function $f$. (The possible exceptions to this are the two
values of $t_1$ defined by the relations $y_1=\pm 2\sqrt{\t(\t+h^2)}/h^2$,
$y_2=-\sinh^{-1} y_1$; when these relations hold, the ratio in \r{newach}
is undefined.) 

\begin{figure}
\centerline{\psfig{figure=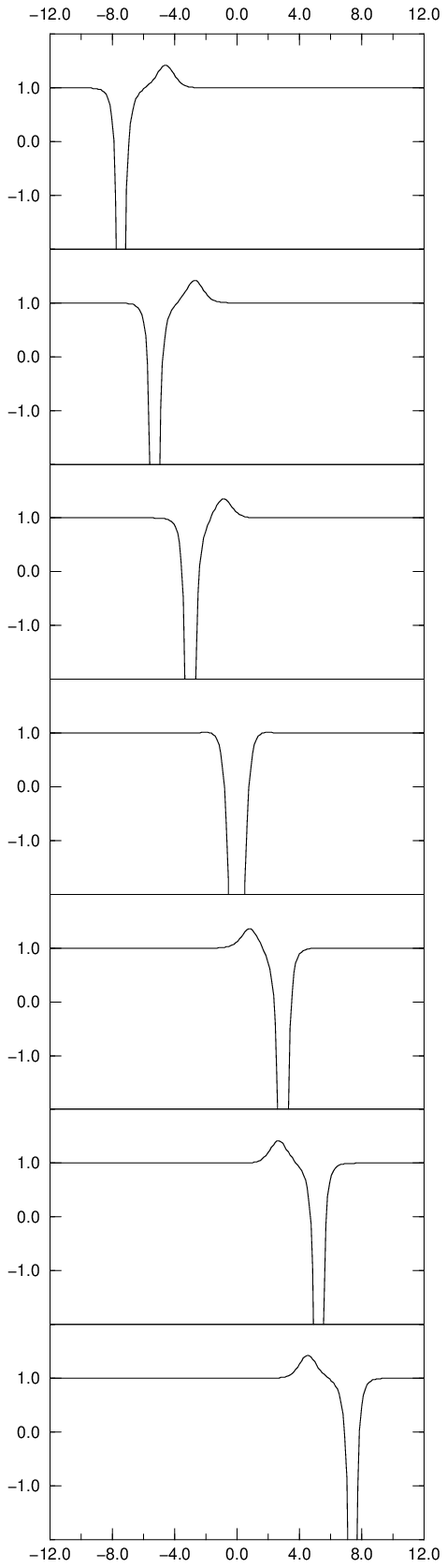,height=6.9in}~~~
            \psfig{figure=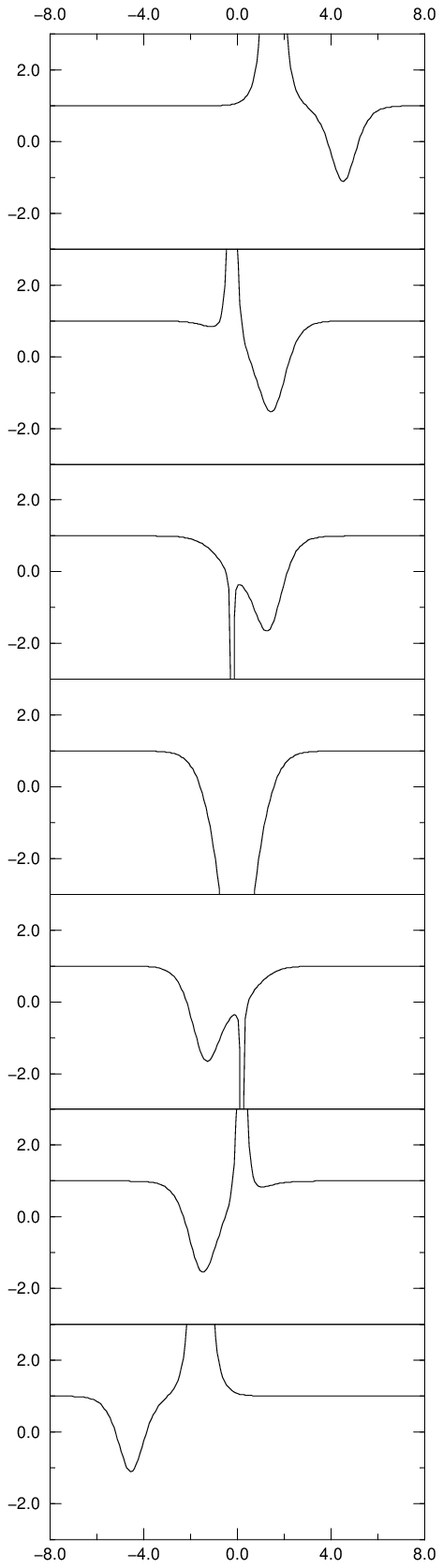,height=6.9in}~~~
            \psfig{figure=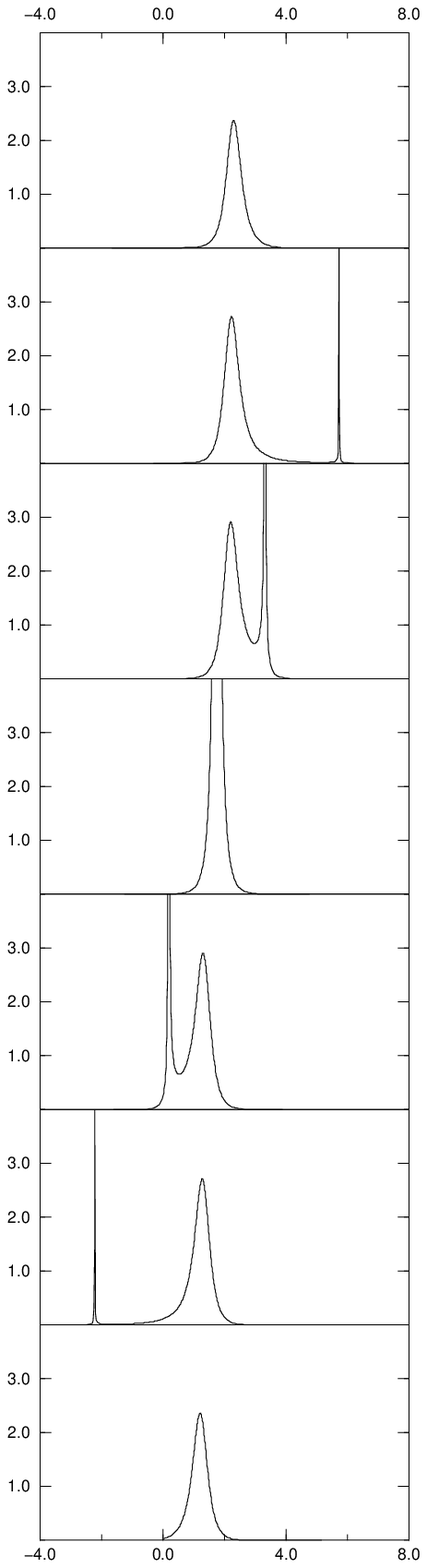,height=6.9in}}
\smallskip
\smallskip
\noindent From left to right:
\smallskip
\caption{The solution \r{newach} of ACH for $h=1$, $\t=-2$, $K_2=K_3=0$.
Plots of $p$ against $t_0$ for $t_1=-3,-2,-1,0,1,2,3$ (from top to bottom).}
\caption{The solution \r{newach} of ACH for $h=1$, $\t=1$, $K_2=K_3=0$.
Plots of $p$ against $t_0$ for $t_1=-3,-0.5,-0.4,0,0.4,0.5,3$ 
(from top to bottom).}
\caption{The solution of CH arising from the middle $t_0$ range 
(the range with $p<0$) of the  solution of ACH of figure 6.
Plots of $u$ against $x$ for $t= -0.5, -0.4055, -0.37, 0, 0.37, 0.4055, 0.5$  
(from top to bottom). The solution is 
analytic for $\vert t \vert > t_c^2 \approx 0.4058$.}
\end{figure}

Let us look more closely at the solution in the case $\t>0$. As can be seen 
in figure 6, as $t_1$ increases, the singularity of $p$ passes through the 
zeros of $p$. The implications of this for the corresponding solutions of 
CH are quite dramatic. Because of the 2 zeros of $p$ there are 3 corresponding
solutions of CH, but let us focus on the solution corresponding to the $t_0$ 
region between the two zeros of $p$. This solution evidently has the 
remarkable feature that for $t$ below one critical time $t_c^1$ and above 
another critical time $t_c^2$ (determined as explained above), the solution 
for $u=p^2$ is analytic, but for $t_c^1<t<t_c^2$, the solution has a 
singularity. The singularity moves in from $x=+\infty$ and out to $x=-\infty$ 
as $t$ increases between the two critical values. The reason a singularity can
develop and disappear this way is that the dynamics of $u$ is  
driven by the dynamics of $f$ and $f$ blows up at both $\pm\infty$. 
It turns out that for the solution \r{newach} it is possible to analytically 
compute the integrals required to change coordinates back from ACH to CH, 
and in figure 7 the interesting solution of CH just described is illustrated
(though this illustration does not adequately
capture the most important feature, that
for $t<t_c^1$ and $t>t_c^2$ there is {\em no} singularity).
For completeness, I note that for $K_2=K_3=0$ 
the critical times are given by
$$
t_c^2 = - t_c^1 = 
   \frac1{2\t(\t+h^2)}\left( 1 + \frac{\sinh^{-1}z}{z}
   \right), \qquad z=\frac{2\sqrt{\t(\t+h^2)}}{h^2} \ .
$$

\section{The Loop Group Construction for ACH}

\subsection{The ACH and CH Hierarchies}

The key to the map from a loop group to solutions of ACH is the following
trivially-checked property:

\noindent{\em Proposition 6.} The ACH equation \r{ch2} has a zero 
curvature formulation
\be \p_{t_1} Z_0 = \p_{t_0} Z_1 + [Z_1,Z_0] \la{zc1}\ee
where
\bea
Z_0 &=& \pmatrix{ 0 & 1/p \cr p/\l + 1/p & 0 \cr}\ , \\
Z_1 &=& \l \pmatrix{0&1\cr 1&0\cr} + 
    \pmatrix{ -\dot{p}/2p &0\cr -2f & \dot{p}/2p \cr}\ ,
\eea
and $f$ is given by the formula in \r{ch2}. 

\smallskip

\noindent  There is an integrable  hierarchy associated with the
CH equation \c{CH}, so it is natural to investigate whether there is a 
hierarchy  related to the ACH equation. Though I do not intend to explore 
it in full detail here, there is a hierarchy, which can be easily
defined using a zero curvature formulation:

\smallskip 

\noindent{\em Definition.} The $n$-th ACH equation ($n \in{\bf Z}$, 
$n\not=0$) is the zero curvature equation 
\be \p_{t_n} Z_0 = \p_{t_0} Z_n + [Z_n,Z_0] \la{zc2}\ee
where 
\be
Z_0 = \pmatrix{ 0 & 1/p(t_0,t_n) \cr p(t_0,t_n)/\l + 1/p(t_0,t_n) & 0 \cr}\ , 
\ee
and 
\begin{itemize}
\item for $n>0$, $Z_n$ is a polynomial in $\l$ of order $n$, with highest
 degree term $\l^n\pmatrix{0&1\cr 1&0\cr}$ and constant term with vanishing
 $1,2$ entry.
\item for $n<0$, $Z_n$ is a polynomial in $1/\l$ of order $1-n$. with
 highest degree term $\l^{n-1}\pmatrix{0&0\cr p(t_0,t_n)&0\cr}$ and
 constant term proportional to $\pmatrix{0&1\cr 1&0\cr}$. 
\end{itemize}

\smallskip

\noindent It is straightforward to check the consistency of the above 
definition. Of particular interest is the $n=-1$ equation, which, 
writing $r=1/p$,  (and ignoring one constant of integration) becomes 
\be
\p_{t_{-1}} r = \left( \frac14r''-\frac38 \frac{r'^2}{r}-
              \frac12 r^3\right)'\ .
\ee
It is also straightforward to use the above definition to show that
the $n>1$ ACH equations are related by a change of coordinates
to the higher equations in
the CH hierarchy, for which a zero curvature formulation is given in 
\c{me1}. Less straightforward (but nevertheless possible) is to 
show the consistency of all the ACH equations, i.e. that we can look
for functions $p(\ldots,t_{-1},t_0,t_1,\dots)$ simultaneously 
satisfying all the equations in the hierarchy. This calculation is
made unnecessary by the loop group construction that will shortly be given,
which constructs solutions of the entire hierarchy.

\subsection{The Loop Group $G$}

Suppose $\e>0$, and denote by 
${\cal C}_0$ and ${\cal C}_\infty$ respectively the circles $\{\vert \l
\vert =\e\}$ and $\{\vert \l \vert =1/\e\}$ in the Riemann sphere. Write 
${\cal C}={\cal C}_0 \cup {\cal C}_\infty$. The loop group we will need,
which I denote $G$, is the 
group of smooth maps from ${\cal C}$ into $SL(2)$. I denote by $G_+$ 
the subgroup of $G$ of maps which are the boundary values
of analytic maps from $\{\e<\vert \l \vert<1/\e\}$ to $SL(2)$, and by $G_-$
the subgroup of $G$ of maps which are the boundary values
of analytic maps $S(\l)$
from $\{\vert \l \vert<\e\}\cup\{\vert \l\vert > 1/\e\}$
to $SL(2)$, satisfying the boundary conditions
\be S(0)=\pmatrix{1/\a & 0\cr \b & \a\cr}\ , \qquad
    S(\infty)=\pmatrix{\sqrt{1+\g^2} & \g \cr \g & \sqrt{1+\g^2}\cr}, 
\la{bcs}\ee
for some $\a,\b,\g$ (with $\a\not=0$).
(It is straightforward to check these conditions do define a group).
Throughout I identify $SL(2)$ with its fundamental
representation. The key property of $G$ we shall use is that a dense open
subset of elements $U\in G$ (given a certain natural topology)
can be factorized in the form
\be U=S^{-1}Y \ee
with $S\in G_-$ and $Y\in G_+$ (the so-called ``Birkhoff factorization''
property).

The corresponding splitting of the Lie algebra ${\cal G}$ of $G$ is
described as follows. An element $v\in{\cal G}$ has Fourier decompositions
on both the circles ${\cal C}_0$ and ${\cal C}_\infty$, i.e. we can write
\bea
v &=&  \sum_{n=-\infty}^{\infty} a_n\l^n \qquad \vert\l\vert=\e \nonumber\\
  &=&  \sum_{n=-\infty}^{\infty} b_n\l^n \qquad \vert\l\vert=1/\e\ ,
   \la{serexp}
\eea
where the coefficients $a_n,b_n$ are in the Lie algebra $sl(2)$. 
Consider the terms containing negative powers of $\l$ in the series valid
on $\vert\l\vert=\e$. Since the series $\sum_{n=1}^{\infty} a_{-n}\l^{-n}$
converges for $\vert\l\vert^{-1}=1/\e$, it converges absolutely for 
$\vert\l^{-1}\vert<1/\e$, i.e. for $\vert\l\vert>\e$, 
defining an analytic function there. Similarly the 
series $\sum_{n=1}^{\infty} b_n\l^{n}$ converges absolutely for 
$\vert\l\vert<1/\e$, defining
an analytic function there. And thus, for arbitrary $t\in sl(2)$ (we will
fix $t$ shortly),
$$ 
v_+=\sum_{n=1}^{\infty} a_{-n}\l^{-n} + t + \sum_{n=1}^{\infty} b_n\l^{n}
$$
defines an analytic function on $\e<\vert\l\vert<1/\e$, and is also convergent
on the boundaries of this region, thus defining an element of the
Lie algebra ${\cal G}_+$ of $G_+$. We have 
\bean
v-v_+ &=& (a_0-t) + \sum_{n=1}^{\infty} (a_n-b_n)\l^n 
           \qquad ~~~~~\vert\l\vert=\e\\
      &=& (b_0-t) + \sum_{n=1}^{\infty} (b_{-n}-a_{-n})\l^{-n}
           \qquad \vert\l\vert=1/\e\ .
\eean
Thus, irrespective of $t$, $v-v_+$ is the boundary value of a function
analytic in $\{\vert\l\vert<\e\}\cup\{\vert\l\vert>1/\e\}$. We choose $t$
so that $v_-=v-v_+$ is in the Lie algebra ${\cal G}_-$ of $G_-$; from 
\r{bcs}, the extra conditions we need to satisfy are 
$$ (v-v_+)(\l=0)=\pmatrix{1-a &0\cr b&1+a\cr}\ , \qquad
    (v-v_+)(\l=\infty)=\pmatrix{0&c \cr c&0\cr}\ , $$
for some $a,b,c$, and a brief
calculation shows that this can be done by taking
$$ t = \pmatrix{(b_0)_{11}&(a_0)_{12}\cr 
   (b_0)_{21}-(b_0)_{12}+(a_0)_{12} & (b_0)_{22}\cr}\ .  $$
To summarize, we have shown the following:

\smallskip

\noindent{\em Proposition 7.} 
For all $v\in{\cal G}$, there exists a unique way to write $v=v_++v_-$ 
with $v_+\in{\cal G}_+$ and $v_-\in{\cal G}_-$. If $v$ has series 
expansions as in \r{serexp}, then
\bean
v_+&=& \sum_{n=1}^{\infty} a_{-n}\l^{-n} + 
   \pmatrix{(b_0)_{11}&(a_0)_{12}\cr 
   (b_0)_{21}-(b_0)_{12}+(a_0)_{12} & (b_0)_{22}\cr} +
       \sum_{n=1}^{\infty} b_n\l^{n} \qquad \e\le\vert\l\vert\le1/\e
   \\
v_-&=& \pmatrix{ (a_0)_{11}-(b_0)_{11} & 0 \cr
                 (b_0)_{12}-(b_0)_{21} & (a_0)_{22}-(b_0)_{22} \cr}
    + \sum_{n=1}^{\infty} (a_n-b_n)\l^n 
           \qquad\qquad ~~~~~~~~~\vert\l\vert\le\e\\
   &=& \pmatrix{0&(b_0)_{12}-(a_0)_{12} \cr
                (b_0)_{12}-(a_0)_{12}&0 \cr}
    + \sum_{n=1}^{\infty} (b_{-n}-a_{-n})\l^{-n}
           \qquad\qquad ~~~~\vert\l\vert\ge1/\e
\eean

\subsection{The Map from $G$ to Solutions of the ACH Hierarchy}

\smallskip

\noindent{\em Proposition 8.} There exists a natural map from the 
loop group $G$ to solutions (possibly with singularities) of the ACH 
hierarchy. This map descends to a map from the coset space $G/G_+$
to solutions of the ACH hierarchy.

\smallskip

\noindent{\em Proof.} The proof I give of this follows the description 
of the Segal-Wilson map given in \c{me2}, which in turn follows the proof
of a similar result for the KP hierarchy given by Mulase \c{Mulase}.
Similar ideas appear in \c{HSS}.

Let ${\cal M}$ denote the infinite dimensional affine manifold with
coordinates $\ldots t_{-1},t_0,t_1,\ldots$, and define a ${\cal G}_+$
valued one-form $\O$ on ${\cal M}$ by
\be \O=\sum_{n=-\infty}^\infty \l^n\pmatrix{0&1\cr 1+\frac1{\l}&0\cr}
        dt_n\ .  \ee
Since evidently $d\O=\O\wedge\O=0$, the differential system
\be dU(t)=\O U(t)\ , \la{Ueq}\ee
where $U$ is a $G$-valued function on ${\cal M}$, is Frobenius integrable,
with general solution 
\be U(t) = M U(0)\ ,  \ee
where 
\bean 
M &=& \exp\left(\sum_{n=-\infty}^\infty \l^n
         \pmatrix{0&1\cr1+\frac1{\l}&0\cr}t_n\right) \\
  &=& \cosh\left(z\sqrt{1+\frac1{\l}}\right) I 
    + \sinh\left(z\sqrt{1+\frac1{\l}}\right) 
      \pmatrix{0&(1+1/\l)^{-1/2}\cr(1+1/\l)^{1/2}&0\cr}\ ,
\eean
and $z=\sum_{n=-\infty}^{\infty}\l^n t_n$ .
Let $U=S^{-1}Y$, $S\in G_-$, $Y\in G_+$, be the Birkhoff decomposition 
of $U$, as described in section 3.2. Substituting into \r{Ueq} we find 
\be -dS\ S^{-1} + dY\ Y^{-1} = S\O S^{-1}\ ,  \ee
from which it follows that 
\be
dS\ S^{-1} = -(S\O S^{-1})_- \qquad {\rm and} \qquad
dY\ Y^{-1} = (S\O S^{-1})_+\ ,
\la{tp}\ee
where here I am using the notation of proposition 7 for the projections
of an element of ${\cal G}$ to ${\cal G}_+$ and ${\cal G}_-$.
If we write $Z=dY\ Y^{-1}$, then clearly $dZ=Z\wedge Z$, or, writing 
\be Z=\sum_{n=-\infty}^{\infty} Z_n dt_n\ , \ee
the components of $Z$ satisfy the zero curvature equations
\be \partial_{t_m} Z_n - \partial_{t_n} Z_m = [Z_m,Z_n] \la{zcbig}\ee
(c.f. \r{zc1},\r{zc2}). On the other hand, the second equation of \r{tp}
tells us that 
\be Z_n=\left( \l^n S \pmatrix{0&1\cr1+\frac1{\l}&0\cr} S^{-1}\right)_+
\ . \la{Zform}\ee
I will now show that this fixes the form of the matrices
$Z_n$ appearing in the zero curvature equations \r{zcbig} to the 
form of the matrices appearing in the zero curvature formulation of
the ACH hierarchy given in section 3.1. From this it follows at once
that given $U(0)\in G$ to specify a solution of \r{Ueq} we can find an
associated solution of the ACH hierarchy, by computing in turn $U$, $S$
(or $Y$) and then $Z=(S\O S^{-1})_+=dY\ Y^{-1}$. This is the natural map
of the proposition. 

To show that \r{Zform} correctly fixes the form of the $Z_n$,
consider first what we know about $S$. $S$ is an element of $G_-$ and
hence has expansions
\bean
S &=& \sum_{n=0}^{\infty} S_n \l^n\ ,~ \qquad 
    S_0 = \pmatrix{1/\a &0\cr\b&\a\cr}\ ,~~~~~~~~~
   ~~~~~~~ \qquad \vert\l\vert \le \e \\ 
  &=& \sum_{n=0}^{\infty} \tilde{S}_n \l^{-n}\ , \qquad 
  \tilde{S}_0 = \pmatrix{\sqrt{1+\g^2} &\g\cr\g&\sqrt{1+\g^2}\cr}\ , 
        \qquad \vert\l\vert \ge 1/\e\ . 
\eean
To see the content of \r{Zform}, we use these formulae to expand 
$ \l^n S \pmatrix{0&1\cr1+\frac1{\l}&0\cr} S^{-1}$ in Laurent series 
valid in $0<\vert\l\vert\le\e$ and $1/\e\le\vert\l\vert<\infty$, and 
then use the projection formula of proposition 7. For example, for $n=0$,
we have, for $0<\vert\l\vert\le\e$ :
\bean
S \pmatrix{0&1\cr1+\frac1{\l}&0\cr} S^{-1}
&=& (S_0+\l S_1)S \pmatrix{0&1\cr1+\frac1{\l}&0\cr} (S_0 + \l S_1)^{-1} 
     + O(\l) \\
&=& \frac1{\l} S_0 \pmatrix{0&0\cr1&0\cr} S_0^{-1} 
  + S_0 \pmatrix{0&1\cr1&0\cr} S_0^{-1} 
  + \left[S_1S_0^{-1}, S_0 \pmatrix{0&0\cr1&0\cr} S_0^{-1} \right]
  + O(\l) \\
&=& \pmatrix{ O(1) & 1/\a^2 + O(\l) \cr
            \a^2/\l + O(1) & O(1) \cr}\ ,
\eean
where here $O(1)$ denotes only non-negative powers of $\l$, and 
for $1/\e\le\vert\l\vert<\infty$
\bean
S \pmatrix{0&1\cr1+\frac1{\l}&0\cr} S^{-1}
&=& \tilde{S}_0 \pmatrix{0&1\cr1&0\cr} \tilde{S}_0^{-1} + O(1/\l) \\
&=& \pmatrix{0&1\cr1&0\cr}  + O(1/\l) \ .
\eean
Using the projection formula we obtain 
\be
Z_0 = \left( S \pmatrix{0&1\cr1+\frac1{\l}&0\cr} S^{-1} \right )_+
    = \pmatrix{ 0  & 1/\a^2  \cr
            \a^2/\l + 1/a^2 & 0 \cr}\ ,
\ee
of the required form with $p=\a^2$. The required results for $Z_n$, 
$n\not=0$, follow in a similar manner; for $n>0$ the expansion for 
$0<\vert\l\vert\le\e$ gives no contribution to the projection, and for
$n<0$ the expansion for $1/\e\le\vert\l\vert<\infty$ gives no 
contribution. 

To conclude the proof, I note that the solutions generated of the ACH
hierarchy may have singularities because the Birkhoff decomposition 
is only possible for open, dense subset of $G$. And, the reason the map
descends to the coset $G/G_+$ is that if we multiply $U(0)$ on the 
right by $g\in G_+$, then $U(t)$ and $Y$ get similarly multiplied,
but $S$ is left unchanged, and therefore so is the solution of ACH.

\section{The Derivation of BTs of ACH}

As explained in \c{me2} for the case of the KdV equation, BTs for ACH
are associated with simple automorphisms of the loop group. 
The relevant automorphisms do not preserve
the fibration of $G$ over $G/G_+$, so a single solution of ACH, 
corresponding to a  $G_+$ coset in $G$, will typically 
get mapped by a BT into a family of solutions, corresponding to a family
of cosets. 

The aim in this section is to outline how the automorphsims
\bea
U(0) &\rightarrow& 
    \sqrt{\frac{\l-\t}{\l+1}}
    \pmatrix{0&1\cr 1+1/\l&0\cr}U(0)\pmatrix{0 & \l/(\l-\t)\cr 1&0\cr}
    \la{aut1}\\
U(0) &\rightarrow& U(0) \left(I + \frac{M}{\l-\t}\right), \qquad
      M=\pmatrix{0&0\cr1&0\cr}\la{aut2}
\eea
give rise, respectively, to the BTs of propositions 3 and 5. This
is simply an exercise in Birkhoff factorization. We will need 
the first terms in the expansion of $Y$ around $\l=\t$, $\t\not=0$.
Without loss of generality (since we have not specified $\e$ in the 
definition of $G$), we assume $\t$ is in the region of analyticity of
$Y$, in which case we can write
$$
Y(\l,t)=Y_0(t)\left[ I + Y_1(t)(\l-\t) + O(\l-\t)^2 \right]\ ,
$$
$$
Y_0(t) = \pmatrix{A(t) & B(t) \cr C(t) & D(t) \cr}\ ,\qquad
Y_1(t) = \pmatrix{a(t) & b(t) \cr c(t) & -a(t) \cr}\ .
$$
($Y_0,Y_1$ of course have $\t$ dependence, but we treat $\t$ as fixed.)
Substituting these expansions into the equations $\p_{t_0}Y=Z_0 Y$ and
$\p_{t_1}Y=Z_1 Y$, and using the forms of $Z_0,Z_1$ given in 
proposition 6 (expanded around $\l=\t$), we find relations between the
fields $A,B,C,D,a,b,c$ and the field $p$; amongst these relations
are the equations
\bean
B' &=& \frac{D}{p} \qquad\qquad~~~~~~
\dot{B} = -\frac{\dot{p}}{2p}B+\t D\\
D' &=& B\left(\frac{p}{\t} + \frac1{p} \right)\qquad
\dot{D} = (\t-2f)B+\frac{\dot{p}}{2p}D \\
b' &=& \frac{pB^2}{\t^2}\qquad\qquad~~~~
\dot{b}=D^2-B^2\ .
\eean
Note that if we eliminate $D$ from this system and write $\tau=\t(1+b)$
we recover the system of equations \r{BT2e1} and \r{BT2e2}. 
$A$ and $C$ obey a set of equation identical to that for $B$ and $D$.

\smallskip

\noindent{\em Derivation of the First BT.} Under the transformation 
\r{aut1}, we have 
\bean 
U(t) &\rightarrow&
    \sqrt{\frac{\l-\t}{\l+1}}
    \pmatrix{0&1\cr 1+1/\l&0\cr}U(t)\pmatrix{0 & \l/(\l-\t)\cr 1&0\cr}  \\
&=& \sqrt{\frac{\l-\t}{\l+1}}
    \pmatrix{0&1\cr 1+1/\l&0\cr}S^{-1} 
    \pmatrix{ -C/A & 1 \cr 1-\t/\l & 0 \cr}^{-1} \cdot
    \pmatrix{ -C/A & 1 \cr 1-\t/\l & 0 \cr}
    Y\pmatrix{0 & \l/(\l-\t)\cr 1&0\cr}  \ ,
\eean
where in the last line, I have factored $U$, and inserted the product
of a certain matrix and its inverse, chosen so that the final expression
is the product of a function analytic in $\{\l<\e\}\cup\{\l>1/\e\}$
and another function analytic in $\{\e<\l<1/\e\}$. (This can be directly
checked; the boundary condition obeyed by $S$ at $\l=0$ must be used.)
This does not quite complete the Birkhoff factorization; it is necessary
to insert a further matrix ---  independent of $\l$ ---
and its inverse in order to make sure the new $S$ satisfies the required
boundary conditions. The calculation is arduous, and I omit the details;
the final result is 
\bean
S &\rightarrow& \frac1{\sqrt{A^2-C^2}}
    \pmatrix{A & 0\cr -C & (A^2-C^2)/A\cr}
    \pmatrix{ -C/A & 1 \cr 1-\t/\l & 0 \cr} 
    \sqrt{\frac{\l+1}{\l-\t}} \ S
    \pmatrix{ 0 & \l/(\l+1) \cr 1&0\cr} \\
Y &\rightarrow& \frac1{\sqrt{A^2-C^2}}
    \pmatrix{A & 0\cr -C & (A^2-C^2)/A\cr}
    \pmatrix{ -C/A & 1 \cr 1-\t/\l & 0 \cr}
    Y\pmatrix{0 & \l/(\l-\t)\cr 1&0\cr}\ .
\eean
Examining the behavior of $S(0)$ we see that the induced transformation
on $\a=1/(S(0))_{11}$ is
$$ \a \rightarrow \frac{\sqrt{\t(C^2-A^2)}}{A}\frac1{\a}\ , $$
and so (since $p=\a^2$),
$$ p \rightarrow \left(\frac{C^2}{A^2}-1\right)\frac{\t}{p}\ . $$
As mentioned above, $A$ and $C$ obey the same equations as 
$B$ and $D$, so we can eliminate $C$ to write the transformation
$$ p \rightarrow \left(\frac{p^2A'^2}{A^2}-1\right)\frac{\t}{p}
       = \t p\left(\frac{A'^2}{A^2}-\frac1{p^2} \right)\ . $$
To make contact with the form in which I have given the BT in section 2,
two more manipulations are necessary. First, note that $s=p\t A'/A$
satisfies equation \r{bt1}, and in terms of this the transformation
becomes simply
$$ p \rightarrow \frac{s^2-\t^2}{\t p}
       = p-s' . $$
Second, it is necessary to use the $t_1$ evolution equation for $A$
to check that $s$ satisfies \r{bt2}; this is completely straightforward.

It just remains to confirm that BTs associated with different values
of $\t$ commute. At first this appears not to be the case. If we denote the 
automorphism \r{aut1} of $G$ by $f(\t)$, then it is simple to check
that the effect of first applying $f(\t_1)$ and then $f(\t_2)$ to $U(0)$
is 
$$ U(0) \rightarrow U(0) \pmatrix{\sqrt{\frac{\l-\t_2}{\l-\t_1}}&0\cr
                          0&\sqrt{\frac{\l-\t_1}{\l-\t_2}}\cr}\ , $$
indicating that order is important. The source of the noncommutativity 
is that in general
$$ \pmatrix{0& \frac{\l}{\l-\t_1} \cr 1&0 \cr}
   \pmatrix{0& \frac{\l}{\l-\t_2} \cr 1&0 \cr} \not=
   \pmatrix{0& \frac{\l}{\l-\t_2} \cr 1&0 \cr}
   \pmatrix{0& \frac{\l}{\l-\t_1} \cr 1&0 \cr}\ . $$
But the BT corresponds to the action of the automorphism on an whole
$G_+$ coset in $G$, and when this is taken into account commutativity is
restored. This can be seen from the simple identity
$$ \pmatrix{0& \frac{\l}{\l-\t_1} \cr 1&0 \cr}
   \pmatrix{0& \frac{\l}{\l-\t_2} \cr 1&0 \cr} =
   \pmatrix{0&1\cr -1&0\cr}
   \pmatrix{0& \frac{\l}{\l-\t_2} \cr 1&0 \cr}
   \pmatrix{0& \frac{\l}{\l-\t_1} \cr 1&0 \cr}
   \pmatrix{0&-1\cr 1&0\cr}\  $$
(by inserting $G_+$ elements --- in fact constant $SL(2)$ matrices 
in this case --- we can make the necessary matrices commute).

\smallskip

\noindent{\em Derivation of the Second BT.} Under the transformation
\r{aut2} we have
\bean 
U(t) &\rightarrow& U(t)\left(I + \frac{M}{\l-\t}\right)  \\
&=& S^{-1} \left(I + \frac{N}{\l-\t}\right)^{-1} \cdot
    \left(I + \frac{N}{\l-\t}\right)
    Y\left(I + \frac{M}{\l-\t}\right)  \ ,
\eean
where once again I have inserted a matrix and  its inverse to make
the last expression the product of functions analytic in appropriate
regions. The correct choice of $N$, which works for any $M$ such 
that $M^2=0$, is
\be N = - Y_0M(I+Y_1M)^{-1}Y_0^{-1}\  \ee
(this satisfies $N^2=0$ and $NY_0M=0$).
Again, this does not complete the Birkhoff factorization, and we need
to insert a constant matrix and its inverse to restore the boundary
condition for $S$ at $\l=0$. We thus get the transformation 
\bean
S &\rightarrow& \pmatrix{ \sqrt{1+h^2} & h\cr h & \sqrt{1+h^2} \cr}
  \left(I + \frac{N}{\l-\t}\right) S \\
Y &\rightarrow& \pmatrix{ \sqrt{1+h^2} & h\cr h & \sqrt{1+h^2} \cr}
  \left(I + \frac{N}{\l-\t}\right)
    Y\left(I + \frac{M}{\l-\t}\right)\ ,
\eean 
where 
\be h = \frac{N_{12}}{\sqrt{(\t-N_{22})^2-N_{12}^2}}\ . \ee
The resulting transformation for $p$ is given by
\be p \rightarrow p\left[\left(1-\frac{N_{22}}{\t}\right)^2
                        -\left(\frac{N_{12}}{\t}\right)^2\right]\ . \ee
So far all the formulas given have been for an arbitrary choice of $M$
in \r{aut2}, with just the proviso that $M^2=0$. For the particular
choice of $M$ indicated we find $N_{12}=B^2/(1+b)$ and 
$N_{22}=BD/(1+b)$. Contact is made with the presentation of the 
BT in section 2 via the system of equations for $B,D,b$ given above.
For this second BT, commutativity for different values of $\t$ is
trivial.

\section*{Acknowledgments}
It is a pleasure to acknowledge hospitality at the University of 
Minneapolis, where some of this work was done, and interesting 
conversations with Yi Li, Peter Olver, Philip Rosenau and David Sattinger.

\end{document}